\documentclass[runningheads,a4paper]{llncs}

\usepackage{hyperref}
\usepackage{color}
\usepackage{amssymb}
\usepackage{amstext}
\usepackage{graphicx}
\usepackage{subfigure}
\usepackage{url,xspace}
\newcommand{\keywords}[1]{\par\addvspace\baselineskip
\noindent\keywordname\enspace\ignorespaces#1}

\newtheorem{model}[theorem]{CWC Modelling}

\newcommand{\qqop}[1]{\mathrel{\makebox[2em]{$#1$}}}

\newcommand{\agr}{\quad\big|\quad}

\newcommand{\AT}{\mathcal{A}}
\newcommand{\LT}{\mathcal{L}}

\newcommand{\LeftPat}{P}

\newcommand{\RightPat}{O}

\newcommand{\srewrites}[1]{\stackrel{#1}{\longmapsto}}

\newcommand{\into}{\ensuremath{\,\rfloor}\,}

\newcommand{\conc}{\;\,}
\newcommand{\emptyseq}{\bullet}

\newcommand{\mapstoDesug}{\mapsto}

\newcommand{\red}{\mapstoDesug}

\newcommand{\ov}[1]{\overline{#1}}

\begin{document}

\mainmatter  

\title{Ecological Modelling with the\\Calculus of Wrapped Compartments\thanks{\scriptsize This work has been partially sponsored by the BioBITs Project (\emph{Converging Technologies 2007}, area:
Biotechnology-ICT), Regione Piemonte. Angelo Troina has visited the Instituto de Ecolog\'{i}a at UTPL (Loja, Ecuador) within the Prometeo program founded by SENESCYT.
A preliminary version of this paper has been presented in~\cite{CMC13}.}}

\titlerunning{Ecological Modelling with the Calculus of Wrapped Compartments}

%
%
\author{Pablo Ram\'{o}n \inst{1} \and  Angelo Troina \inst{2}}
\authorrunning{P.~Ram\'{o}n and A.~Troina}

\institute{Instituto de Ecolog\'{i}a, Universidad Tecnica Particular de Loja, Ecuador
\and
Dipartimento di Informatica, Universit\`{a} di Torino, Italy
}

%
%

\toctitle{Lecture Notes in Computer Science}
\tocauthor{Authors' Instructions}
\maketitle

\begin{abstract}

The Calculus of Wrapped Compartments is a framework based on stochastic multiset rewriting in a compartmentalised setting originally developed for the modelling and analysis of biological interactions.
In this paper, we propose to use this calculus for the description of ecological systems and we provide the modelling guidelines to encode within the calculus some of the main interactions leading ecosystems evolution.
As a case study, we model the distribution of height of \emph{Croton wagneri}, a shrub constituting the endemic predominant species of the dry ecosystem in southern Ecuador. In particular, we consider the plant at different altitude gradients (i.e. at different temperature conditions), to study how it adapts under the effects of global climate change.
\end{abstract}

\section{Introduction}
\label{intro}
Answers to ecological questions could rarely be formulated as general laws: ecologists deal with \emph{in situ} methods and experiments which cannot be controlled in a precise way since the phenomena observed operate on much larger scales (in time and space) than man can effectively study. Actually, to carry on ecological analyses, there is the need of a ``macroscope''! 

Theoretical and Computational Ecology, the scientific disciplines devoted to the study of ecological systems using theoretical methodologies together with empirical data, could be considered as a fundamental component of such a macroscope. Within these disciplines, quantitative analysis, conceptual description techniques, mathematical models,  and computational simulations are used to understand the  fundamental biological conditions and processes that affect populations dynamics (given the underlying assumption that phenomena observable across species and ecological environments are generated by common, mechanistic processes)~\cite{Pie77}.

Ecological models can be deterministic or stochastic~\cite{Bol08}. Given an initial system, deterministic simulations always evolve in the same way, producing a unique output~\cite{SM90}. Deterministic methods give a picture of the average, expected behaviour of a system, but do not incorporate random fluctuations. On the other hand, stochastic models allow to describe the random perturbations that may affect natural living systems, in particular when considering small populations evolving at slow interactions. Actually, while deterministic models are approximations of the real systems they describe, stochastic models, at the price of an higher computational cost, can describe exact scenarios.

A model in the Calculus of Wrapped Compartments (CWC for short) consists of a term, representing a (biological or ecological) system and a set of rewrite rules which model the transformations determining the system's evolution~\cite{preQAPL2010,CDDGGT_TCSB11}. Terms are defined from a set of atomic elements via an operator of compartment construction. Each compartment is labelled with a nominal type which identifies the set of rewrite rules that may be applied into it.
The CWC framework is based on a stochastic semantics and models an exact scenario able to capture the stochastic fluctuations that can arise in the system.

The calculus has been extensively used to model real biological scenarios, in particular related to the AM-symbiosis~\cite{CDDGGT_TCSB11,spatial_COMPMOD11}.\footnote{Arbuscular  Mycorrhiza (AM) is a class of fungi constituting a vital mutualistic interaction for terrestrial ecosystems. More than 48\% of land plants actually rely on mycorrhizal relationships to get inorganic compounds, trace elements, and resistance to several kinds of pathogens.} An hybrid semantics for CWC, combining stochastic transitions with deterministic steps, modelled by Ordinary Differential Equations, has been proposed in~\cite{HCWC_mecbic10,CWCTCS}.

While the calculus has been originally developed to deal with biomolecular interactions and cellular communications, it appears to be particularly well suited also to model and analyse interactions in ecology. In particular, we present in this paper some modelling guidelines to describe, within CWC, some of the main common features and models used to represent ecological interactions and population dynamics. A few generalising examples illustrate the abstract effectiveness of the application of CWC to ecological modelling.

As a real case study, we model the distribution of height of \emph{Croton wagneri}, a shrub in the dry ecosystem of southern Ecuador, and investigate how it could adapt to global climate change.

\subsection{Motivation and Methodology}

At the beginning of the second half of the twentieth century, when ecology was still a young science and mathematical models for ecological systems were in their infancy, Elton~\cite{Elt58}, acknowledging the influence of Lotka \cite{Lot25} and Volterra~\cite{Vol26}, wrote: ``Being mathematicians, they did not attempt to contemplate a whole food--chain with all the complications of five stages. They took two: a predator and its prey''.
Nowadays, in the era of computational ecological modelling, deterministic systems based on ordinary differential equations for two variables, or even a whole food chain, appear like simple idealisations quite distant from the real complexity of nature. Predator--prey interactions are now considered  as ``consumer--resource'' interactions embedded within the large ecological networks that underlie biodiversity. Lotka--Volterra equations and their many descendants assume that individuals within a system are well mixed and interact at mean population abundances. They are mean-field equations that use the mass--action law to describe the dynamics of interacting populations, and ignore both the scale of individual interactions and their spatial distribution. However, because ecological systems are typically nonlinear, they often cannot be solved analytically, and, in order to obtain sensible results, nonlinear, stochastic computational techniques must be used.

The formal framework to be used as the modelling core of this project should thus be able to manage several features which are typical of ecological systems. Namely, complex ecological systems are multilevel, they follow non linear, stochastic dynamics and involve a distributed spatial organisation.\\

\noindent \textbf{Multilevel Modelling.} The role of the computational methodology used to model and simulate ecological systems is to address questions on the relationship between systems dynamics at different temporal, spatial, and organisational (or structural) scales. In particular, it is important to address the variability at small, local scales and its effects on the dynamics of the aggregated quantities measured at large, global scales~\cite{Pas05}.\\

\noindent \textbf{Stochastic Modelling.} Ecological models can be deterministic or stochastic~\cite{Bol08}. Given an initial system, deterministic simulations always evolve in the same way, producing a unique output~\cite{SM90}. Deterministic methods give a picture of the average, expected behaviour of a system, but do not incorporate random fluctuations. On the other hand, stochastic models allow to describe the random perturbations that may affect natural living systems, in particular when considering small populations evolving at slow interactions. Actually, while deterministic models are approximations of the real systems they describe, stochastic models, at the price of an higher computational cost, can describe exact scenarios. Stochastic models, such as interacting particle systems, can also help us examine new approaches for scaling up individual--based dynamics.\footnote{Note that the impact of stochastic factors and the corresponding level of a system's uncertainty are much higher in Ecology than in other natural sciences. Common sources of uncertainty are, e.g., the poor accuracy of ecological data and their transient nature. Noise that is inevitably present in ecosystems can significantly change the properties of an ecological model and this fundamental uncertainty affects the accuracy of ecological data \cite{PP12}.}\\

\noindent \textbf{Spatial Modelling.} The impact of space tends to make the population dynamics significantly more complicated compared with its non--spatial counterpart and to bring new and bigger challenges to simulations. Formal models dealing explicitly with spatial coordinates are able to depict more precise localities and \emph{ecological niches}, describing, for example, how organisms or populations respond to the distribution of resources and competitors~\cite{LB98}.

\noindent \textbf{The Calculus of Wrapped Compartments.} While the Calculus of Wrapped Compartments has been originally developed to deal with biomolecular interactions and cellular communications, it appears to be particularly well suited also to model and analyse interactions in ecology.
The Calculus of Wrapped Compartments satisfies the main requirements addressed in the previous section. Namely, CWC is able to model and simulate: (i) multilevel systems, (ii) stochastic dynamics, (iii) explicit spatial systems.

The compartment operator of the calculus can be used to describe the topological organisation of a systems. It also allows to deal with multilevel systems by defining different set of rules for different compartments, reflecting the interactions taking place at the different levels of the system.

The evolution of a system described in CWC follows a stochastic simulation model defined by incorporating a collision-based framework along the lines of the one presented by Gillespie in~\cite{G77}, which is, \emph{de facto}, the standard way to model quantitative aspects of biological systems. The basic idea of Gillespie's algorithm is that a rate is associated with each considered reaction. This rate is used as the parameter of an exponential probability distribution modelling the time needed for the reaction to take place. In the standard approach, the reaction \emph{propensity} is obtained by multiplying the rate of the reaction by the number of possible combinations of reactants in the compartment in which the reaction takes place, modelling the law of mass action.

A spatial extension of CWC has been proposed in~\cite{BCCDSST11}, incorporating a two--dimensional spatial description of the elements in the system through axial coordinates and special rules for the movement of system components in space. The spatial extension of the calculus can be generalised to deal with spaces defined in more than two dimensions.

\paragraph{Summary}
In Section~\ref{cwc} we recall the syntax and semantics of CWC. In Section~\ref{ecolog} we present some of the characteristic processes leading ecosystems evolution and show how to encode them in CWC. In Section~\ref{case} we model the distribution of \emph{Croton wagneri}, a shrub in the dry ecosystem of southern Ecuador, and investigate how it could adapt to global climate change. Finally, in Section~\ref{conclu} we draw our conclusions and survey some related work.

\section{The Calculus of Wrapped Compartments}
\label{cwc}

The Calculus of Wrapped Compartments (CWC)
(see~\cite{preQAPL2010,HCWC_mecbic10,CWCTCS}) is based on a nested structure of compartments delimited by wraps with specific proprieties.

\subsection{Term Syntax}
\label{CWC_formalism - syntax}

Let $\AT$ be a set of  \emph{atomic elements} (\emph{atoms} for
short), ranged over by $a$, $b$, ..., and  $\LT$ a set of \emph{compartment types} represented as \emph{labels} ranged over by $\ell,\ell',\ell_1,\ldots$

\begin{definition}[CWC terms]
A CWC \emph{term} is a multiset $\ov{t}$ of  \emph{simple terms} $t$ defined by the following grammar:
$$
t::= \quad a \agr (\overline{a}\into \overline{t'})^\ell
$$
\end{definition}

A simple term is either an atom or a compartment consisting of a \emph{wrap} (represented by the multiset of atoms $\overline{a}$), a \emph{content} (represented by the term
$\overline{t'}$) and a \emph{type} (represented by the label $\ell$).
Multisets are identified modulo permutations of their elements.
The notation $n*t$ denotes $n$ occurrences of the simple term $t$. We denote an empty term with $\emptyseq$.

In applications to ecology, atoms can be used to describe the individuals of different species and compartments can be used to distinguish different ecosystems, habitats or ecological niches. Compartment wraps can be used to model geographical boundaries or abiotic components (like radiations, climate, atmospheric or soil conditions, etc.). In evolutionary ecology, individuals can also be described as compartments, showing characteristic features of their \emph{phenotype} in the wrap and keeping their \emph{genotype} (or particular \emph{alleles} of interest) in the compartment content.

An example of CWC term is $20\!*\!a \conc 12*b \conc (c \conc d \into 6*e \conc
4*f)^\ell$ representing a multiset (denoted by listing its elements separated by a space) consisting of 20 occurrences of $a$, 12 occurrence
of $b$ (e.g. 32 individuals of two different species) and an $\ell$-type compartment $(c \conc d \into 6*e \conc 4*f)^\ell$ which, in turn, consists of a wrap (a boundary) with two
atoms $c$ and $d$ (e.g. two abiotic factors) on its surface, and containing 6 occurrences of the atom $e$ and 4 occurrences of the atom $f$ (e.g. 10 individuals of two other species). Compartments can be nested as in the term $(a \conc b \conc c \into (d \conc e \into f)^{\ell'} \conc g \conc h)^\ell$.

\subsection{Rewrite Rules}
\label{CWC-RR}

System transformations are defined by rewrite rules, defined by resorting to CWC terms that may contain variables.

\begin{definition}[Patterns and Open terms] \emph{Simple patterns} $\LeftPat$ and \emph{simple open terms} $\RightPat$ are given by the following grammar:
$$
\begin{array}{lcl}
   \LeftPat  & \;\qqop{::=}\; & a \agr (\overline{a} \conc x \into \overline{\LeftPat}\conc X)^\ell \\
   \RightPat & \;\qqop{::=}\; & a \agr (\ov{q} \into \ov{\RightPat})^\ell   \agr X\\
   q         & \;\qqop{::=}\; & a \agr x
\end{array}
$$
where $\overline{a}$ is a multiset of atoms,  $\overline{\LeftPat}$ is a \emph{pattern} (i.e., a, possibly empty, multiset of simple patterns), $x$ is a \emph{wrap
variable} (can be instantiated by a multiset of atoms), $X$ is a \emph{content variable}
 (can be instantiated by a CWC term), $\ov{q}$ is a multiset of atoms and wrap variables and $\overline{\RightPat}$ is an \emph{open term} (i.e., a, possibly empty, multiset of simple open terms).
\end{definition}

We will use patterns as the l.h.s. components of a rewrite rule and open terms as the r.h.s. components of a rewrite rule.
Patterns are intended to match, via substitution of variables, with ground terms (containing no variables).
Note that we force \emph{exactly} one variable to occur in each compartment content and wrap of our patterns.
This prevents ambiguities in the instantiations needed to match a given compartment.\footnote{
 The linearity condition, in biological terms, corresponds to excluding that a transformation can depend on the presence of two (or more) identical (and generic) components in different compartments (see also~\cite{OP11}).}

\begin{definition}[Rewrite rules]
A \emph{rewrite rule} is a triple $(\ell, \ov{\LeftPat},\ov{\RightPat})$, denoted by $\ell:  \overline{\LeftPat}  \srewrites{}    \ov{\RightPat}$, where the pattern $\ov{\LeftPat}$ and the open term $\ov{\RightPat}$ are such that
the variables occurring in $\ov{\RightPat}$ are a subset of the variables occurring in $\ov{\LeftPat}$.
\end{definition}

The rewrite rule  $\ell:  \ov{\LeftPat} \red \ov{\RightPat}$ can be applied to any compartment of type $\ell$ with $\overline{\LeftPat}$ in its content (that will be rewritten with $\overline{\RightPat}$). Namely, the application of $\ell:  \ov{\LeftPat} \red \ov{\RightPat}$ to term $\ov{t}$ is performed in the following way:
\begin{enumerate}
\item find in $\ov{t}$ (if it
exists) a compartment of type $\ell$ with content $\ov{t'}$ and a substitution $\sigma$ of variables by ground terms such that $\ov{t'} = \sigma( \ov{\LeftPat} \conc
X)$;\footnote{The implicit (distinguished) variable $X$ matches with all the remaining part of the compartment content.}
\item replace in $\ov{t}$ the
subterm $\ov{t'}$ with $\sigma(\ov{\RightPat}\conc X)$.
\end{enumerate}
%

For instance, the rewrite rule
  $\ell: a \conc b \red c$
means that in compartments of type $\ell$ an occurrence of $a \conc b$  can be replaced by $c$.
We write  $\ov{t} \red \ov{t'}$ to denote a \emph{reduction} obtained by applying a rewrite rule to $\ov{t}$ resulting to $\ov{t'}$.

While a rewrite rule does not change the label $\ell$ of the compartment where it is applied, it may change the labels of the compartments occurring in its content. For instance, the rewrite rule  $\ell: (a \conc x \into X)^{\ell_1} \red (a \conc x \into X)^{\ell_2}$ means that, if contained in a compartment of type $\ell$, a compartment of type $\ell_1$ containing an $a$ on its wrap can be changed to type $\ell_2$.


\paragraph{CWC Models.} For uniformity reasons we assume that the whole system is always represented by a term consisting of a single (top level) compartment with distinguished label $\top$ and empty wrap, i.e., any system is represented by a term of the shape $(\emptyseq \into \ov{t})^{\top}$, which, for simplicity, will be written as $\ov{t}$.
Note that while an infinite set of terms and rewrite rules can be defined from the syntactic definitions in this section, a \emph{CWC model} consists of an initial system $(\emptyseq \into \ov{t})^{\top}$ and a finite set of rewrite rules ${\cal R}$.

\subsection{Stochastic Simulation}\label{SECT:STO_SEM}

A stochastic simulation model for ecological systems can be defined by incorporating a collision-based framework along the lines of the one presented by Gillespie in \cite{G77}, which is, \emph{de facto}, the standard way to model quantitative aspects of biological systems. The basic idea of Gillespie's algorithm is that a rate is associated with each considered reaction which is used as the parameter of an exponential probability distribution modelling the time needed for the reaction to take place. In the standard approach the reaction \emph{propensity} is obtained by multiplying the rate of the reaction by the number of possible combinations of reactants in the compartment in which the reaction takes place, modelling the law of mass action.

Stochastic rewrite rules are thus enriched with a rate $k$ (notation $\ell:  \ov{\LeftPat} \srewrites{k} \ov{\RightPat} $).
Evaluating the propensity of the stochastic rewrite rule $R=
\ell: a \conc b  \srewrites{k} c$ within the term $\ov{t}=a\conc a \conc a \conc b \conc b$, contained in the compartment $u=(\emptyseq \into \ov{t})^\ell$, we must consider the number of the possible combinations of reactants of
the form $a\conc b$ in $\ov{t}$. Since each occurrence of $a$ can react with each occurrence of $b$, this number is $3\cdot 2$, and the propensity of $R$ within $u$ is $k\cdot 6$.
A detailed method to compute the number of combinations of reactants can be found in~\cite{preQAPL2010}.

The stochastic simulation algorithm produces
essentially a \emph{Continuous Time Markov Chain} (CTMC).
Given a term $\ov{t}$, a set ${\cal R}$ of rewrite rules, a global
time $\delta$ and all the reductions $e_1,\ldots,e_M$ applicable in all the different compartments of
$\ov{t}$ with propensities $r_1,\ldots,r_M$, Gillespie's ``direct method'' determines:

\begin{itemize}
\item The exponential probability distribution (with parameter $r = \sum_{i=1}^M
r_i$) of the time $\tau$ after which the next reduction will occur;
\item The probability $r_i/r$ that the reduction occurring at time
  $\delta+\tau$ will be $e_i$.\footnote{Reductions are applied in a sequential way, one at each step.}
\end{itemize}

The CWC simulator~\cite{HCWC_SIM} is a tool under development at the Computer Science Department of the Turin University, based on Gillespie's direct method algorithm~\cite{G77}. It treats CWC models with different rating semantics (law of mass action, Michaelis-Menten kinetics, Hill equation) and it can run independent stochastic simulations over CWC models, featuring deep parallel optimizations for multi-core platforms on the top of FastFlow~\cite{fastflow:web}. It also performs online analysis by a modular statistical framework~\cite{ACDDTT_PDP11,ACDDSSTT_HIBB11}. 

\section{Modelling Ecological Systems in CWC}
\label{ecolog}
Computational Ecology is a field devoted to the quantitative description and analysis of ecological systems using empirical data, mathematical models (including statistical models), and computational technology. While the different components of this interdisciplinary field of research are not new, there is a new emphasis on the integrated treatment of the area. This emphasis is amplified by the expansion of our local, national, and international computational infrastructure, coupled with the heightened social awareness of ecological and environmental issues and its effects on research funding.

We advocate a convergence between computer and life sciences. This emerging paradigm moves to a system level understanding of life, where unpredictable,
complex behaviour show up. We claim that computer science will greatly contribute to a
better understanding of the behaviour of ecological systems. We plan to develop models, languages and tools for describing, analysing and
implementing \emph{in silico} ecological systems, as an additional contribution of Information Technology to
those typical research areas in current Computational Ecology, such as (i) storing, organising and
retrieving large amounts of ecological data or (ii) visual modelling techniques for scientific visualisation of multi--dimensional, computer--generated scenes that can be used to express empirical data.

More in detail, we use our formal framework for modelling and studying the
behaviour of living systems. Our starting point is that ecological systems are conveniently described as
entities that change their state because of the occurrence of biotic and abiotic interactions, giving
rise to some observable behaviour. We thus adhere to the view of living
systems as biological computing units.

In this section we present some of the characteristic features leading the evolution of ecological systems, and we show how to encode them within CWC.

\subsection{Population Dynamics}

Models of population dynamics describe the changes in the size and composition of populations.

The \emph{exponential growth model} is a common mathematical model for population dynamics, where, using $r$ to represent the pro-capita growth rate of a population of size $N$, the change of the population is proportional to the size of the already existing population:
$$
\frac{dN}{dt} = r \cdot N
$$

\begin{model}[Exponential Growth Model] We can encode within CWC the exponential growth model with rate $r$ using a stochastic rewrite rule describing a reproduction event for a single individual at the given rate. Namely, given a population of species $a$ living in an environment modelled by a compartment with label $\ell$, the following CWC rule encodes the exponential growth model:
$$
\ell: a  \srewrites{r} a \conc a
$$
Counting the number of possible reactants, the growth rate of the overall population is automatically obtained by the stochastic semantics underlying CWC.
\end{model}

A \emph{metapopulation}\footnote{The term metapopulation was coined by Richard Levins in 1970. In Levins' own words, it consists of ``a population of populations''~\cite{Lev69}.} is a group of populations of the same species distributed in different patches\footnote{A patch is a relatively homogeneous area differing from its surroundings.} and interacting at some level.
Thus, a metapopulation consists of several distinct populations and areas of suitable habitat.

Individual populations may tend to reach extinction as a consequence of demographic stochasticity (fluctuations in population size due to random demographic events); the smaller the population, the more prone it is to extinction. A metapopulation, as a whole, is often more stable: immigrants from one population (experiencing, e.g., a population boom) are likely to re-colonize the patches left open by the extinction of other populations. Also, by the \emph{rescue effect}, individuals of more dense populations may emigrate towards small populations, rescuing them from extinction.

Populations are affected by births and deaths, by immigrations and emigrations (BIDE model~\cite{Cas01}). The number of individuals at time $t+1$ is given by:
$$N_{t+1}=N_{t}+B+I-D-E$$
where $N_t$ is the number of individuals at time $t$ and, between time $t$ and $t+1$, $B$ is the number of births, $I$ is the number of immigrations, $D$ is the number of deaths and $E$ is the number of emigrations.
Conditions triggering migration could be: climate, food availability or mating~\cite{DD07}.

\begin{model}[BIDE model] We can encode within CWC the BIDE model for a compartment of type $\ell$ using stochastic rewrite rules describing the given events with their respective rates $r$, $i$, $d$, $e$:
$$
\begin{array}{lr}
\ell: a  \srewrites{r} a \conc a & \quad \text{(birth)}\\
\top: a \conc (x \into X)^\ell \srewrites{i} (x \into a \conc X)^\ell  & \quad \text{(immigration)}\\
\ell: a \srewrites{d} \emptyseq & \quad \text{(death)}\\
\top: (x \into a \conc X)^\ell  \srewrites{e} a \conc (x \into X)^\ell & \quad \text{(emigration)}
\end{array}
$$
Starting from a population of $N_t$ individuals at time $t$, the number $N_{t+1}$ of individuals at time  $t+1$ is computed by successive simulation steps of the stochastic algorithm. The race conditions computed according to the propensities of the given rules assure that all of the BIDE events are correctly taken into account.
\end{model}

\begin{example} Immigration and extinction are key components of island biogeography. We model a metapopulation of species $a$ in a context of 5 different patches: 4 of which are relatively close, e.g. different ecological regions within a small continent, the last one is far away and difficult to reach, e.g. an island. The continental patches are modelled as CWC compartments of type $\ell_c$, the island is modelled as a compartment of type $\ell_i$. Births, deaths and migrations in the continental patches are modelled by the following CWC rules:
$$
\begin{array}{c}
\ell_c: a  \srewrites{0.005} a \conc a \quad \quad
\ell_c: a \srewrites{0.005} \emptyseq\\
\top: (x \into a \conc X)^{\ell_c}  \srewrites{0.01} a \conc (x \into X)^{\ell_c} \quad \quad
\top: a \conc (x \into X)^{\ell_c} \srewrites{0.5} (x \into a \conc X)^{\ell_c}
\end{array}
$$
These rates are drawn considering days as time unites and an average of life expectancy and reproduction time for the individuals of the species $a$ of 200 days ($\frac{1}{0.005}$). For the modelling of real case studies, these rates could be estimated from data collected \emph{in situ} by tagging individuals.\footnote{In the remaining examples we will omit a detailed time description.} In this model, when an individual emigrates from its previous patch it moves to the top-level compartment from where it may reach one of the close continental patches (might also be the old one) or start a journey through the sea (modelled as a rewrite rule putting the individual on the wrapping of the island compartment):
$$
\top: a \conc (x \into X)^{\ell_i} \srewrites{0.2} (x\conc a \into  X)^{\ell_i}\\
$$
Crossing the ocean is a long and difficult task and individuals trying it will probably die during the cruise; the luckiest ones, however, might actually reach the island, where they could eventually benefit of a better life expectancy for them and their descendants:
$$
\begin{array}{c}
\top: \conc (x\conc a \into X)^{\ell_i} \srewrites{0.333} (x \into  X)^{\ell_i} \quad \quad
\top: \conc (x\conc a \into X)^{\ell_i} \srewrites{0.0005} (x \into a\conc X)^{\ell_i}\\
\ell_i: a  \srewrites{0.007} a \conc a \quad \quad
\ell_i: a \srewrites{0.003} \emptyseq
\end{array}
$$
Considering the initial system modelled by the CWC term:
$$
\ov{t}= (\emptyseq \into 30*a)^{\ell_c} \conc (\emptyseq \into 30*a)^{\ell_c} \conc (\emptyseq \into 30*a)^{\ell_c} \conc (\emptyseq \into 30*a)^{\ell_c} \conc (\emptyseq \into \emptyseq)^{\ell_i}
$$
we can simulate the possible evolutions of the overall diffusion of individuals of species $a$ in the different patches. Notice that, on average, one over $\frac{0.333}{0.0005}$ individuals that try the ocean journey, actually reach the island.
In Figure~\ref{FigMP} we show the result of a simulation plotting the number of individuals in the different patches in a time range of approximatively 10 years. Note how, in the final part of the simulation, empty patches get recolonised. In this particular simulation, also, an exponential growth begins after the colonisation of the island.
The full CWC model describing this example can be found at: \url{http://www.di.unito.it/~troina/cmc13/metapopulation.cwc}.
\begin{figure}
\centering
\includegraphics[height=60mm]{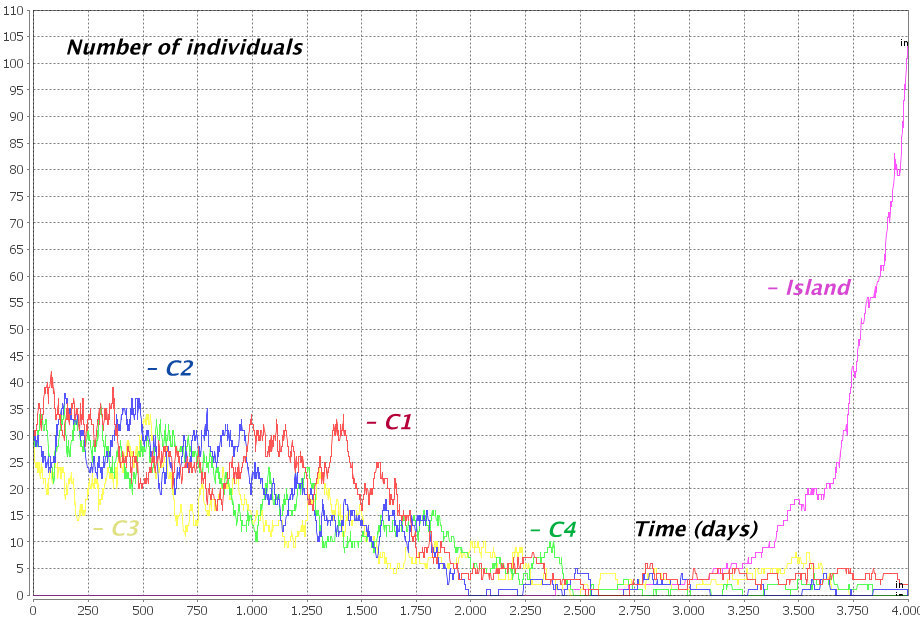}
\caption{\label{FigMP} Metapopulation dynamics.}
\end{figure}

\end{example}

In ecology, using $r$ to represent the pro-capita growth rate of a population and $K$ the \emph{carrying capacity} of the hosting environment,\footnote{I.e., the population size at equilibrium.} $r/K$ selection theory~\cite{Pia70} describes a selective pressure driving populations evolution through the \emph{logistic model}~\cite{Verhulst}:
$$
    \frac{dN}{dt} = r\cdot N \cdot \left(1 - \frac{N}{K}\right)
$$
where $N$ represents the number of individuals in the population.

\begin{model}[Logistic Model] The logistic model with growth rate $r$ and carrying capacity $K$, for an environment modelled by a compartment with label $\ell$, can be encoded within CWC using two stochastic rewrite rules describing (i) a reproduction event for a single individual at the given rate and (ii) a death event modelled by a fight between two individuals at a rate that is inversely proportional to the carrying capacity:
$$
\begin{array}{l}
\ell: a  \srewrites{r} a \conc a\\
\ell: a \conc a \srewrites{\frac{2\cdot r}{K-1}} a
\end{array}
$$
If $N$ is the number of individuals of species $a$, the number of possible reactants for the first rule is $N$ and the number of possible reactants for the second rule is, in the exact stochastic model, ${N \choose 2}=\frac{N\cdot(N-1)}{2}$, i.e. the number of distinct pairs of individuals of species $a$. Multiplying this values by the respective rates we get the propensities of the two rules and can compute the value of $N$ when the equilibrium is reached (i.e., when the propensities of the two rules are equal): $r\cdot N = \frac{2 \cdot r}{K-1}\cdot \frac{N\cdot (N-1)}{2}$, that is when $N=0$ or $N=K$.
\end{model}

For a given species, this model allows to describe different growth rates and carrying capacities in different ecological regions. Identifying a CWC compartment type (through its label) with an ecological region, we can define rules describing the growth rate and carrying capacity for each region of interest.

Species showing a high growth rate are selected by the $r$ factor, they usually exploit low-crowded environments and produce many offspring, each of which has a relatively low probability of surviving to adulthood. By contrast, $K$-selected species adapt to densities close to the carrying capacity, tend to strongly compete in high-crowded environments and produce fewer offspring, each of which has a relatively high probability of surviving to adulthood.

\begin{example}
There is little, or no advantage at all, in evolving traits that permit successful competition with other organisms in an environment that is very likely to change rapidly, often in disruptive ways. Unstable environments thus favour species that reproduce quickly ($r$-selected species).
Characteristic traits of $r$-selected species include: high fecundity, small body, early reproduction and short generation time.
Stable environments, by contrast, favour the ability to compete successfully for limited resources ($K$-selected species). Characteristic traits of $K$-selected species include: large body size, long life expectancy, production of fewer offspring (usually requiring extensive parental care until maturity).
We consider individuals of two species, $a$ and $b$.
Individuals of species $a$ are modelled with an higher growth rate with respect to individuals of species $b$ ($r_a>r_b$). Carrying capacity for species $a$ is, instead, lower than the carrying capacity for species $b$ ($K_a<K_b$).
The following CWC rules describe the $r/K$ selection model for $r_a=5$, $r_b=0.00125$, $K_a=100$ and $K_b=1000$:
$$
\begin{array}{c}
\ell: a  \srewrites{5} a \conc a \quad \quad
\ell: b  \srewrites{0.00125} b \conc b\\
\ell: a \conc a \srewrites{0.1} a \quad \quad
\ell: b \conc b \srewrites{0.0000025} b
\end{array}
$$
We might consider a disruptive event occurring on average every 4000 years with the rule:
$$
\top: (x \into X)^{\ell} \srewrites{0.00025} (x \into a \conc b)^{\ell}
$$
devastating the whole content of the compartment (modelled with the variable $X$) and just leaving one individual of each species.
In Figure~\ref{FigrK} we show a 10000 years simulation for an initial system containing just one individual for each species. Notice how individuals of species $b$ are disadvantaged with respect to individuals of species $a$ who reach the carrying capacity very soon. A curve showing the growth of individuals of species $b$ in a stable (non disruptive) environment is also shown. The full CWC model describing this example can be found at: \url{http://www.di.unito.it/~troina/cmc13/rK.cwc}.
\begin{figure}
\centering
\includegraphics[height=50mm]{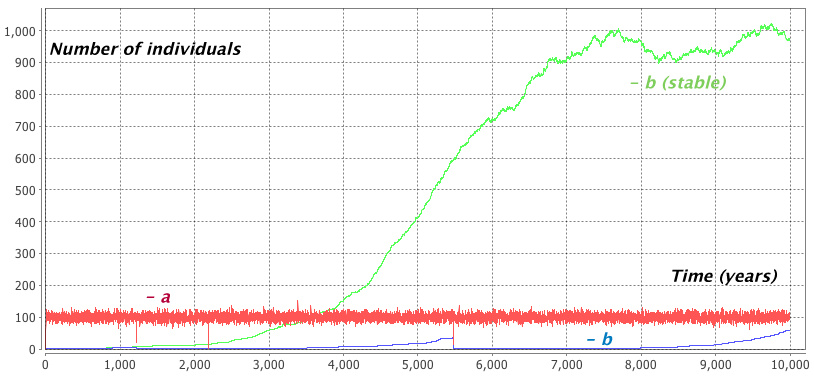}
\caption{\label{FigrK} r/K selection in a disruptive environment.}
\end{figure}
\end{example}

\subsection{Competition and Mutualism}

In ecology, \emph{competition} is a contest for resources between organisms: animals, e.g., compete for water supplies, food, mates, and other biological resources. In the long term period, competition among individuals of the same species (\emph{intraspecific competition}) and among individuals of different species (\emph{interspecific competition}) operates as a driving force of adaptation, and, eventually, by natural selection, of evolution. Competition, reducing the fitness of the individuals involved,\footnote{By fitness it is intended the ability of surviving and reproducing. A reduction in the fitness of an individual implies a reduction in the reproductive output. On the opposite side, a fitness benefit implies an improvement in the reproductive output.} has a great potential in altering the structure of populations, communities and the evolution of interacting species. It results in the ultimate survival, and dominance, of the best suited variants of species: species less suited to compete for resources either adapt or die out. We already depicted a form of competition in the context of the logistic model, where individuals of the same species compete for vital space (limited by the carrying capacity $K$).

Quite an apposite force is \emph{mutualism}, contest in which organisms of different species biologically interact in a relationship where each of the individuals involved obtain a fitness benefit. Similar interactions between individuals of the same species are known as \emph{co-operation}. Mutualism belongs to the category of symbiotic relationships, including also \emph{commensalism} (in which one species benefits and the other is neutral, i.e. has no harm nor benefits) and \emph{parasitism} (in which one species benefits at the expense of the other).

The general model for
competition and mutualism between two species $a$ and $b$ is defined by the following equations~\cite{Tak89}:
$$
\begin{array}{l}
   \frac{dN_a}{dt} = \frac{r_a \cdot N_a}{K_a} \cdot \left( K_a - N_a + \alpha_{ab}\cdot N_b \right) \\
   \frac{dN_b}{dt} = \frac{r_b \cdot N_b}{K_b} \cdot \left( K_b - N_b + \alpha_{ba}\cdot N_a \right)
\end{array}
$$
where the $r$ and $K$ factors model the growth rates and the carrying capacities for the two species, and
the $\alpha$ coefficients describe the nature of the relationship between the two species: if $\alpha_{ij}$ is negative, species $N_j$ has negative effects on species $N_i$ (i.e., by competing or preying it), if $\alpha_{ij}$ is positive, species $N_j$ has positive effects on species $N_i$ (i.e., through some kind of mutualistic interaction).

The logistic model, already discussed, is included in the differential equations above. Here we abstract away from it and just focus on the components which describe the effects of competition and mutualism we are now interested in.

\begin{model}[Competition and Mutualism] For a compartment of type $\ell$, we can encode within CWC the model about competition and mutualism for individuals of two species $a$ and $b$ using the following stochastic rewrite rules:
$$
\ell: a \conc b \srewrites{f_a \cdot |\alpha_{ab}|}
\left\{
\begin{array}{lr}
 a\conc a \conc b & \quad \text{if } \alpha_{ab}>0 \\
 b & \quad \text{if } \alpha_{ab}<0
\end{array}
\right.
\quad\quad
\ell: a \conc b \srewrites{f_b \cdot |\alpha_{ba}|}
\left\{
\begin{array}{lr}
 a\conc b \conc b & \quad \text{if } \alpha_{ba}>0 \\
 a & \quad \text{if } \alpha_{ba}<0
\end{array}
\right.
$$
where $f_i=\frac{r_i}{K_i}$ is obtained from the usual growth rate and carrying capacity. The $\alpha$ coefficients are put in absolute value to compute the rate of the rule, their signs affect the right hand part of the rewrite rule.
\end{model}

\begin{example}
Mutualism has driven the evolution of much of the biological diversity we see today, such as flower forms (important to attract mutualistic pollinators) and co-evolution between groups of species~\cite{Tho05}.
We consider two different species of pollinators, $a$ and $b$, and two different species of angiosperms (flowering plants), $c$ and $d$. The two pollinators compete between each other, and so do the angiosperms. Both species of pollinators have a mutualistic relation with both angiosperms, even if $a$ slightly prefers $c$ and $b$ slightly prefers $d$. For each of the species involved we consider the rules for the logistic model and for each pair of species we consider the rules for competition and mutualism. The parameters used for this model are in Table~\ref{TabMC}. So, for example, the mutualistic relations between $a$ and $c$ are expressed by the following CWC rules
$$
\top: a \conc c \srewrites{\frac{r_a}{K_a}\cdot \alpha_{ac}} a \conc a \conc c
\quad \quad \quad
\top: a \conc c \srewrites{\frac{r_c}{K_c}\cdot \alpha_{ca}} a \conc c \conc c
$$
Figure~\ref{FigCM} shows a simulation obtained starting from a system with 100 individuals of species $a$ and $b$ and 20 individuals of species $c$ and $d$. Note the initially balanced competition between pollinators $a$ and $b$. This random fluctuations are resolved by the ``long run'' competition between the angiosperms $c$ and $d$: when $d$ predominates over $c$ it starts favouring the pollinator $b$ that now can win its own competition with pollinator $a$. The model is completely symmetrical: in other runs, a faster casual predominance of a pollinator may lead the evolution of its preferred angiosperm. The CWC model describing this example can be found at: \url{http://www.di.unito.it/~troina/cmc13/compmutu.cwc}.

\begin{table}
\footnotesize
\centering
\begin{tabular}{|c|c|c|c|c|c|c|}
\hline
\textbf{Species} ($i$)	&	$r_i$		&	$K_i$	&	$\alpha_{ai}$	& $\alpha_{bi}$	& $\alpha_{ci}$& $\alpha_{di}$	\\
\hline
$a$								&	0.2			& 1000		& $\bullet$ 			& -1						& +0.03			 & +0.01				\\
\hline
$b$								&  0.2			& 1000		& -1						& $\bullet$			& +0.01			& +0.03				\\
\hline
$c$								&  0.0002		& 200		& +0.25				& +0.1					& $\bullet$ 		& -6						\\
\hline
$d$								&  0.0002		& 200		& +0.1					& +0.25				& -6					 & $\bullet$			\\
\hline
\end{tabular}
\normalsize
\caption{\label{TabMC} Parameters for the model of competition and mutualism.}
\end{table}

\begin{figure}
\centering
\includegraphics[height=50mm]{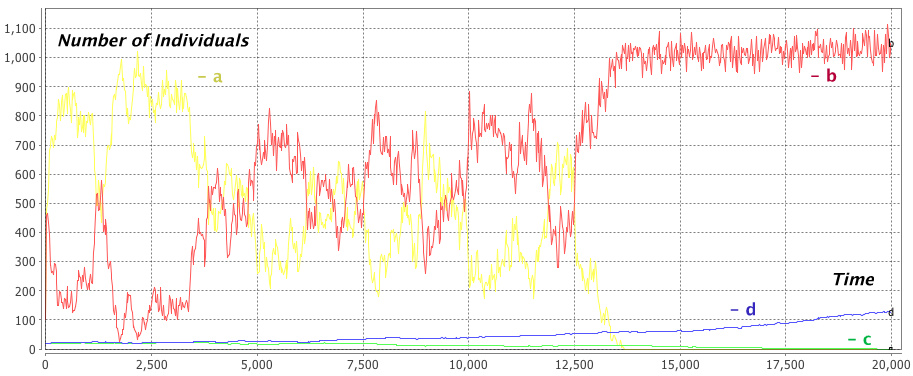}
\caption{\label{FigCM} Competition and Mutualism.}
\end{figure}

\end{example}

\subsection{Trophic Networks}

A \emph{food web} is a network mapping different species according to their alimentary habits. The edges of the network, called \emph{trophic links}, depict the feeding pathways (``who eats who'') in an ecological community~\cite{Elt27}. At the base of the food web there are autotroph species\footnote{Self-feeding: able to produce complex organic compounds from simple inorganic molecules and light (by \emph{photosynthesis}) or inorganic chemical reactions (\emph{chemosynthesis}).}, also called basal species. A \emph{food chain} is a linear feeding pathway that links monophagous consumers (with only one exiting trophic link) from a top consumer, usually a larger predator, to a basal species. The length of a chain is given by the number of links between the top consumer and the base of the web. The influence that the elements of a food web have on each other determine important features of an ecosystem like the presence of strong interactors (or \emph{keystone species}), the total number of species, and the structure, functionality and stability of the ecological community.

To model quantitatively a trophic link between species $a$ and $b$ (i.e., a particular kind of competition) we might use Lotka-Volterra equations~\cite{Vol26}:
$$
\begin{array}{l}
    \frac{dN_b}{dt} = N_b \cdot (r_b-\alpha \cdot N_a) \\
    \frac{dN_a}{dt} = N_a\cdot (\beta \cdot N_b - d)
\end{array}
$$
where $N_a$ and $N_b$ are the numbers of predators and preys, respectively, $r_b$ is the rate for prey growth, $\alpha$ is the prey mortality rate for per-capita predation, $\beta$ models the efficiency of conversion from prey to predator and $d$ is the mortality rate for predators.

\begin{model}[Trophic Links] Within a compartment of type $\ell$, given a predation mortality $\alpha$ and conversion from prey to predator $\beta$, we can encode in CWC a trophic link between individuals of species $a$ (predator) and $b$ (prey) by the following rules:
$$
\begin{array}{l}
\ell: a \conc b \srewrites{\alpha} a\\
\ell: a \conc b \srewrites{\beta} a \conc a \conc b
\end{array}
$$
Here we omitted the rules for the prey exponential growth (absent predators) and predators exponential death (absent preys). These factors are present in the Lotka-Volterra model between two species, but could be substituted by the effects of other trophic links within the food web. In a more general scenario, a trophic link between species $a$ and $b$ could be expressed condensing the two rules within the single rule:
$$
\ell: a \conc b \srewrites{\gamma} a \conc a
$$
with a rate $\gamma$ modelling both the prey mortality rate and the predator conversion factor.
\end{model}

\begin{example}
Trophic cascades occur when predators in a food web suppress the abundance of their prey, thus limiting the predation of the next lower trophic level. For example, an herbivore species could be considered in an intermediate trophic level between a basal species and an higher predator. Trophic cascades are important for understanding the effects of removing top predators from food webs, as humans have done in many ecosystems through hunting or fishing activities. We consider a three-level food chain between species $a$, $b$ and $c$. The basal species $a$ reproduces with the logistic model, the intermediate species $b$ feeds on $a$, species $c$ predates species $b$:
\footnotesize
$$
\ell: a \srewrites{0.4} a \conc a \qquad
\ell: a \conc a\srewrites{0.0002} a
\qquad \qquad
\ell: a \conc b \srewrites{0.0004} b \conc b \qquad
\ell: b \conc c \srewrites{0.0008} c \conc c
$$
\normalsize
Individuals of species $c$ die naturally, until an hunting species enters the ecosystem. At a rate lower than predation, $b$ may also die naturally (absent predator). An atom $h$ may enter the ecosystem and start hunting individuals of species $c$:
\footnotesize
$$
\ell: c \srewrites{0.52} \emptyseq \qquad
\ell: b \srewrites{0.03} \emptyseq
\qquad \qquad
\top: h \conc (x \into X)^\ell \srewrites{0.003} (x \into X \conc h)^\ell \qquad
\ell: h \conc c \srewrites{0.5} h
$$
\normalsize
Figure~\ref{FigTC} shows a simulation for the initial term $h \conc (\emptyseq \into 1000*a \conc 100*b \conc 10*c )^\ell$. When the hunting activity starts, by removing the top predator, a top-down cascade destroys the whole community. The CWC model describing this example can be found at: \url{http://www.di.unito.it/~troina/cmc13/trophic.cwc}.
\begin{figure}
\centering
\includegraphics[height=50mm]{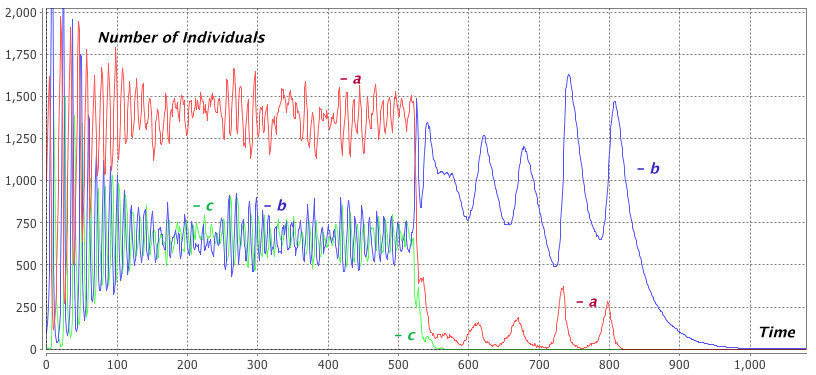}
\caption{\label{FigTC} A Throphic Cascade.}
\end{figure}
\end{example}

\section{An application: \emph{Croton wagneri} and Climate Change}
\label{case}
The knowledge of the relationships between the various attributes that define natural ecosystems (composition, structure, biotic and abiotic interactions, etc.) is crucial to improve our understanding of their functioning and dynamics. Much in the same way, this knowledge could allow us to evaluate and predict the ecological impacts of global change and to establish the appropriate measures to preserve the intrinsic characteristics of the ecological systems minimising the environmental impact.

Dry ecosystems are characterised by the presence of discontinuous vegetation that may reflect less than 60\% of the available landscape. The main pattern in arid ecosystems is a vegetation mosaic composed of patches and clear sites.
Dry ecosystems constitute an ideal model to analyse the relationships between the attributes of the ecosystem and its functioning at different levels of organisation. This becomes particularly important for the crucial need, for this kind of ecosystem, to deal with environmental problems like the loss of biodiversity and climate change \cite{ReySS02}. Morphological and functional changes shown by each species of these ecosystems are generally very different. As a consequence, the continuously adapting structure of the community will depend on the plasticity of each species in response to the major environmental changes. In the last decades, changes in the morphology of a species have been investigated from a functional perspective. This kind of analysis becomes particularly relevant because, for example, the existence of plant species in environments with extreme climatic conditions may depend not only on the ability of the plant to structurally specialise but also on its capacity to adapt metabolically. Another key factor of dry ecosystems is the possibility to study special patterns of the vegetation distribution to understand whether the plant communities show competitive or mutualistic relations. In particular, in dry (infertile) ecosystems, when competitive relations predominate, the spatial distribution of the vegetation tends to be uniform (with a quite regular distance between the plants); when mutualistic relations predominate, the vegetation tends to form clusters of plants; if none of these relations is observed, the spatial distribution of the vegetation is more random \cite{WM04}.

The study site is located in a dry scrub in the south of Ecuador ($03^\circ 58' 29''$ S, $01^\circ 25' 22''$ W) near the Catamayo Valley, with altitude ranging from 1400m to 1900m over the sea level. Floristically, in this site we can find typical species of xerophytic areas (about 107 different species and 41 botanical families). The seasonality of the area directly affects the species richness: about the 50\% of the species reported in the study site emerge only in the rainy season. Most species are shrubs although there are at least 12 species of trees with widely scattered individuals, at least 50\% of the species are herbs.
The average temperature is $20^\circ$ C with an annual rainfall around 600 mm, the most of the precipitation occurs between December and March, the remaining months present a water deficit with values between 200 and 600 mm of rain annually.
Generally, this area is composed by clay, rocky and sandy soils~\cite{AKS05}.

In~\cite{Gen93} about 1300 different species belonging to the dry ecosystems in Northwest South America have been identified. For this study we will focus on data available for the species \emph{Croton wagneri M\"{u}ll. Arg.}, belonging to the Euphorbiaceae family. This species, particularly widespread in tropical regions, can be identified by the combination of latex, alternate simple leaves, a pair of glands at the apex of the petiole, and the presence of stipules. \emph{C. wagneri} is the dominant endemic shrub in the dry scrub of Ecuador and has been listed as Near Threatened (NT) in the Red Book of Endemic Plants of Ecuador~\cite{VPLJ00}. This kind of shrub could be considered as a nurse species\footnote{A nurse plant is one with an established canopy, beneath which germination and survival are more likely due to increased shade, soil moisture, and nutrients.} and is particularly important for its ability to maintain the physical structure of the landscape and for its contribution to the functioning of the ecosystem (observing a marked mosaic pattern of patches having a relatively high biomass dispersed in a matrix of poor soil vegetation)~\cite{Gut01}.

In the study area, 16 plots have been installed along four levels of altitude gradients (1400m, 1550m, 1700m and 1900m): two 30mx30m plots per gradient in plane terrain and two 30mx30m plots per gradient in a slope surface (with slope greater than $10^\circ$). The data collection survey consisted in enumerating all of the \emph{C. wagneri} shrubs in the 16 plots: the spatial location of each individual was registered using a digital laser hypsometer. Additionally, plant heights were measured directly for each individual and the crown areas were calculated according to the method in~\cite{SACH11}. Weather stations collect data about temperatures and rainfall for each altitude gradient. An extract of data collected from the field can be found at: \url{http://www.di.unito.it/~troina/croton_data_extract.xlsx}.
This data show a morphological response of the shrub to two factors: temperature and terrain slope. A decrease of the plant height is observed at lower temperatures (corresponding to higher altitude gradients), or at higher slopes.

\subsection{The CWC model}

A simulation plot is modelled by a compartment with label $P$. Atoms $g$, representing the plot gradient (one $g$ for each metre of altitude over the level of the sea), describe an abiotic factor put in the compartment wrap.

According to the temperature data collected by the weather stations we correlate the mean temperatures in the different plots with their respective gradients. In the content of a simulation plot, atoms $t$, representing 1$^\circ$C each, model its temperature. Remember that, in this case, the higher the gradient, the lower the temperature. Thus, we model a constant increase of temperature within the simulation plot compartment, controlled by the gradient elements $g$ on its wrap:
$$
\top: (x \into X)^P  \srewrites{1}   (x \into t \conc X)^P \qquad
\top: (g \conc x \into t \conc X)^P  \srewrites{0.000024}   (g \conc x \into  X)^P
$$


Atoms $i$ are also contained within compartments of type $P$, representing the complementary angle of the plot's slope (e.g., $90*i$ for a plane plot or $66*i$ for a 24$^\circ$ slope).

We model \emph{C. wagneri} as a CWC compartment with label $c$. Its observed trait, namely the plant height, is specified by atomic elements $h$ (representing one mm each) on the compartment wrap.

To model the shrub heights distribution within a parcel, we consider the plant in two different states: a ``young'' and an ``adult'' state. Atomic elements $y$ and $a$ are exclusively, and uniquely, present within the plant compartment in such a way that the shrub height increases only when the shrub is in the young state ($y$ in its content). The following rules describe (i) the passage of the plant from $y$ to $a$ state with a rate corresponding to a 1 year average value, and (ii) the growth of the plant, affected by temperature and slope, with a rate estimated to fit the field collected data:
$$
c: y  \srewrites{0.00274}  a \qquad \qquad
P: t \conc i \conc (x \into y \conc X)^{c}  \srewrites{0.000718}   t \conc i \conc (x \conc h \into y \conc X)^c
$$

\subsection{Simulation results}
Now we have a model to describe the distribution of  \emph{C. wagneri} height using as parameters the plot's gradient ($n*g$) and slope ($m*i$). Since we do not model explicitly interactions that might occur between \emph{C. wagneri} individuals, we consider plots containing a single shrub. Carrying on multiple simulations, through the two phase model of the plant growth, after 1500 time units (here represented as days), we get a snapshot of the distribution of the shrubs heights within a parcel. The CWC model describing this application can be found at: \url{http://www.di.unito.it/~troina/cmc13/croton.cwc}.

\begin{figure}
\centering
\subfigure[$1400*g$ and $90*i$] {
\includegraphics[height=36mm]{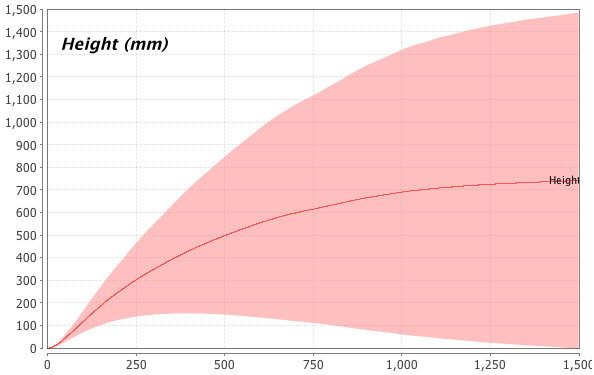}
}
\centering
\subfigure[$1550*g$ and $60*i$] {
\includegraphics[height=36mm]{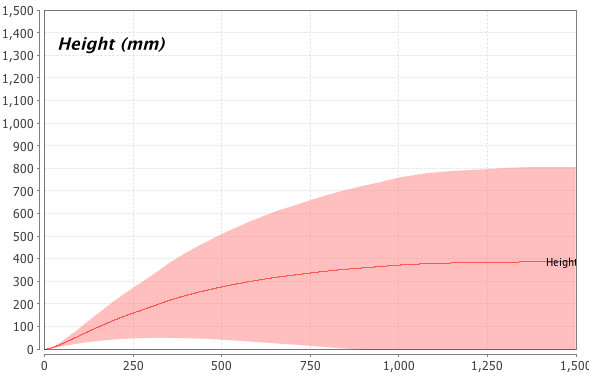}
}
\subfigure[$1700*g$ and $85*i$] {
\includegraphics[height=36mm]{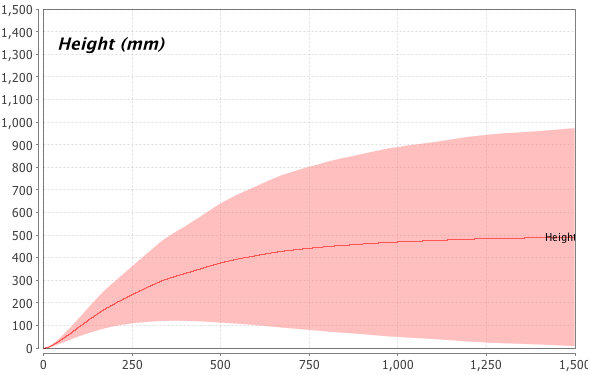}
}
\subfigure[$1900*g$ and $75*i$] {
\includegraphics[height=36mm]{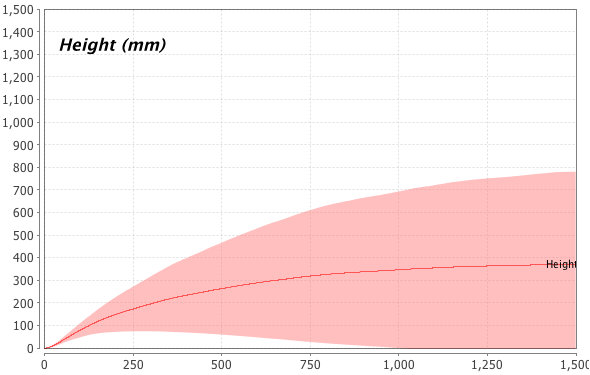}
}

\caption{Deviation of the height of \emph{Croton wagneri} for 100 simulations.}
\label{FigC}
\end{figure}

Each of the graphs in Figure~\ref{FigC} is obtained by plotting the height deviation of 100 simulations with initial term $(n*g \into m*i \conc (\emptyseq \into y)^c)^P$.
The simulations in Figures~\ref{FigC} (a) and (c) reflect the conditions of real plots and the results give a good approximation of the real distribution of plant heights. Figures~\ref{FigC} (b) and (d) are produced considering an higher slope than the ones on the real plots from were the data has been collected. These simulation results can be used for further validation of the model by collecting data on new plots corresponding to the parameters of the simulation.

\begin{figure}
\centering
\subfigure[$12^\circ$C, plain terrain] {
\includegraphics[height=36mm]{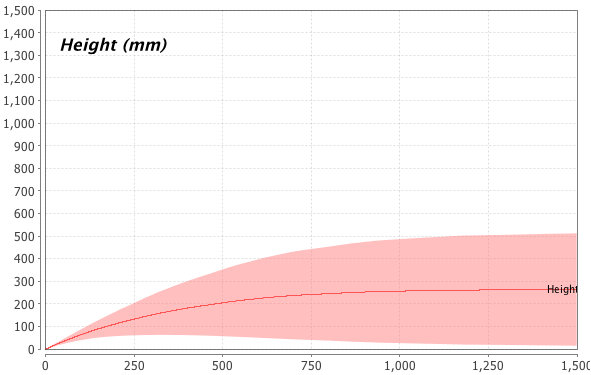}
}
\centering
\subfigure[$10^\circ$C, plain terrain] {
\includegraphics[height=36mm]{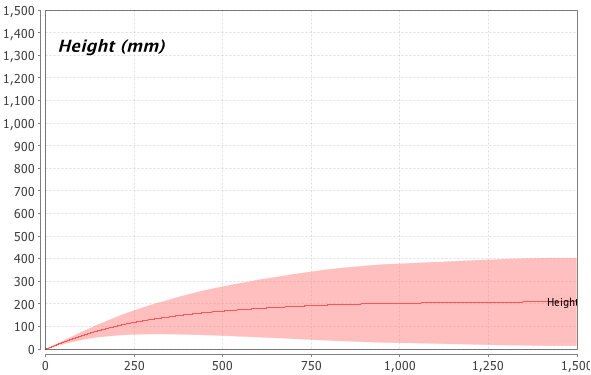}
}

\caption{Deviation of the height of \emph{Croton wagneri} for 100 simulations.}
\label{FigCT}
\end{figure}

If we already trust the validity of our model, we can remove the correlation between the gradient and the temperature, and directly express the latter. Predictions can thus be made about the shrub height at different temperatures, and how it could adapt to global climate change. Figure~\ref{FigCT} shows two possible distributions of the shrub height at lower temperatures (given it will actually survive these more extreme conditions and follow the same trend).

\section{Conclusions and Related Works}
\label{conclu}
The long-term goal of Computational Ecology is the development of methods to predict the response of ecosystems to changes in their physical, chemical and biological components. Computational models, and their ability to understand and predict the biological world, could be used to express the mechanisms governing the structure and function of natural populations, communities, and ecosystems. Until recent times, there was insufficient computational power to run stochastic, individually-based, spatially explicit models. Today, however, some of these techniques could be investigated~\cite{PP12}.

The main objective of Computational Ecology is twofold. First, we aim at discovering new models and theories within
Computer Science inspired by the natural world, and at producing techniques and tools to
deal with much more complex problems than those addressable with
current technology. Second, we plan to provide ecologists with an environment for attacking
problems at a system level, not addressable without using Information Technology. This
environment will provide ecologists with modelling, analysis and simulation tools capable of
handling complex behaviour and of representing emerging properties.

Calculi developed to describe process interactions in a compartmentalised setting are well suited for the description and analysis of the evolution of ecological systems. The topology of the ecosystem can be directly encoded within the nested structure of the compartments. These calculi can be used to represent structured natural processes in a greater detail, when compared to purely numerical analysis. As an example, food webs can give rise to combinatorial interactions resulting in the formation of complex systems with emergent properties (as signalling pathways do in cellular biology), and, in some cases, giving rise to chaotic behaviour.

Also, \emph{ecological niches} describe how organisms or populations respond to the distribution of resources and competitors~\cite{LB98}, and arise as the spatial sectors where organisms and populations tend to distribute, often forming geographical clusters of dominant species. Using the Spatial Calculus of Wrapped Compartments~\cite{BCCDSST11,spatial_COMPMOD11}, we might be able to extend our analysis to the case of ecological niches and show how they  could emerge from complex ecological interactions and then become a fundamental characteristic of an ecosystem.

As a final remark about ecological modelling with a framework based on stochastic rewrite rules, we underline an important compositional feature. How can we test an hypothetical scenario where a grazing species is introduced in the model of our case study? A possibility could be to represent the grazing species with a new CWC atom (e.g. $s$) and then just add the new competitive rules to the previously validated model (e.g. the rule $P: s \conc (h \conc x \into X)^c \srewrites{} s \conc ( x \into X)^c$). Changing in the same sense a model based on ordinary differential equations would, instead, result in a complete new model were all previous equations should be rewritten.

\subsection{Related works}

Computer simulations play
a rather unique role in Ecology, compared with other sciences, and there are several reasons for the extensive growth in
the number of simulation--based studies in ecological
mathematical modelling. In particular, the
problem is that, although a field experiment is a common
research approach in ecology, replicated experiments
under controlled conditions (which is a cornerstone of
all natural sciences providing information for theory
development and validation) are rarely possible because
of the transient nature of the environment: just consider, for instance,
the impossibility to reproduce the same weather
pattern for a repeated experiment in an ecological study. We also mention it here that
large--scale ecological experiments are costly and, in the situation
when consequences are poorly understood, can have
adverse effects on some species, put in danger the biodiversity of the ecosystem,
and may even pose a threat to human well--being.
Capturing the complexity of real systems through tractable
experiments is therefore logistically not feasible.
Mathematical modelling and computer simulations
create a convenient ``virtual environment'' and hence
can provide a valuable supplement, or sometimes even
an alternative, to the field experiment.\\

\noindent \textbf{Ecological Modelling.} It has long been recognised that numerical modelling and computer simulations can be used
as a powerful research tool to understand, and sometimes to predict, the tendencies and
peculiarities in the dynamics of populations and ecosystems. It has been, however, much
less appreciated that the context of modelling and simulations in ecology is essentially different
from those that normally exist in other natural sciences~\cite{PP12}.

Ecology became a quantitative and
theory--based science since the seminal studies by Lotka \cite{Lot25}, Volterra \cite{Vol26} and Gause \cite{Gau34}, who were the first to use mathematical
tools for ecological problems. General principles of ecosystem
organisation were later refined and systematised by
Odum \& Odum~\cite{OO53}, while the mathematical theory was
further developed by Skellam~\cite{Ske51} and Turing~\cite{Tur52}, who
emphasised the importance of the spatial aspect.
The bright mathematical ideas of those seminal works
sparked a huge fire. The last quarter of the twentieth century
saw an outbreak of interest in mathematical ecology
and ecological modelling \cite{WO76,OMWM89,DMW91,MWD93,SLF95,DMW98}. Especially over the last
decade, more and more complicated models have been
developed with a generic target to take into account
the ecological interactions in much detail and hence to
provide an accurate description of ecosystems dynamics.
Owing to their increased complexity, many of the models
had to be solved numerically, a development that was
inspired and made possible by the simultaneous advances in computer
science and technology.

The modelling approaches can be very different in terms of the mathematics used and depending on the goals of the study, and there are several ways to classify them. For instance, there is an apparent difference between statistical models \cite{Cza98} and ``mechanistic'' models \cite{Mur93,OL01}, although simulation--based studies may sometimes include both. Taken from another angle, two qualitatively different modelling streams are rule--based approaches (such as individual-based models and cellular automata) \cite{GR05} and equation--based ones.

Another way to sort out the models used in ecology is to consider the level of complexity involved. Depending on the purposes of the ecological study, there have been two different streams in model building \cite{May74}. In case the purpose is to predict the system's state (with a certain reasonable accuracy), the model is expected to include as many details as possible. This approach is often called \emph{predictive modelling}. The mathematical models arising in this way can be very complicated and analytically intractable in an exact way (in these cases the model can still be partially analysed via a limited number of runs of computer simulations) \cite{PCSKR96,NCO11,Pas05,GW92}. Alternatively, the purpose of the study can be to understand the current features of the ecosystem, e.g. to identify the factors responsible for a population decline or a population outbreak, but not necessarily to predict their development quantitatively. We will call this approach a \emph{conceptual modelling}. In this case, the corresponding models can be pretty simple, even if their exact solutions are still not always possible; therefore, they often have to be solved by simulations as well~\cite{OMWM89,DMW91,SLF95}.
These two streams of theoretical--ecological research can be clearly seen in the literature, even though it is not always straightforward to distinguish between them as sometimes simple models may show a certain predictive power and, on the contrary, complicated ones are used for making a qualitative insight into some subtle issues.\\

\noindent \textbf{Formal Computational Frameworks.}
As P-Systems~\cite{Pau00,Pau02} and the Calculus of Looping Sequences (CLS, for short)~\cite{BMMT07}, the Calculus of Wrapped Compartments is a framework modelling topological compartmentalisation inspired by biological membranes, and with a semantics given in terms of rewrite rules.

CWC has been developed as a simplification of CLS, focusing on stochastic multiset rewriting. The main difference between CWC and CLS consists in the exclusion of the sequence operator, that constructs ordered strings out of the atomic elements of the calculus. While the two calculi keep the same expressiveness, some differences arise on the way systems are described. On the one hand, the Calculus of Looping Sequences allows to define ordered sequences in a more succinct way (for examples when describing sequences of genes in DNA or sequences of amino acids in proteins).\footnote{An ordered sequence can be expressed in CWC as a series of nested compartments, ordered from the outermost compartment to the innermost one.} On the other hand, CWC reflects in a more realistic way the fluid mosaic model of the lipid bilayer (for example in the case of cellular membrane description, where proteins are free to float), and, the addition of compartment labels allows to characterise the properties peculiar to given classes of compartments. Ultimately, focusing on multisets and avoiding to deal explicitly with ordered sequences (and, thus, variables for sequences) strongly simplifies the pattern matching procedure in the development of a simulation tool.

The Calculus of Looping Sequences has been extended with type systems in~\cite{ADT09,DGT09,DGT09b,BDMMT10,BDGT12}. As an application to ecology, stochastic CLS (see~\cite{BMMTT08}) is used in~\cite{BCBMMR10} to model population dynamics.

P-Systems have been proposed as a computational model inspired by biological structures. They are defined as a nesting of membranes in which multisets of objects can react according to pre defined rewrite rules. Maximal-parallelism is the key feature of P-Systems: at each evolution step all rewrite rules, in all membranes, are applied as many times as possible. Such a feature makes P-Systems a very powerful computational model and a versatile instrument to evaluate expressiveness of languages. However, it is not practical to describe stochastic systems with a maximally-parallel evolution: exact stochastic simulations based on race conditions model systems evolutions as a sequence of successive steps, each of which with a particular duration modelled by an exponential probability distribution.

There is a large body of literature about applications of P-Systems to ecological modelling. In~\cite{CCMPPPS11,CCMPPS09,CCPSM09}, P-Systems are enriched with a probabilistic semantics to model different ecological systems in the Catalan Pyrenees. Rules could still be applied in a parallel fashion since reduction durations are not explicitly taken into account. In~\cite{BCPM07,BCPM08,BCPM10}, P-Systems are enriched with a stochastic semantics and used to model metapopulation dynamics. The addition of \emph{mute rules} allows to keep a form of parallelism reducing the maximal consumption of objects.

While all these calculi allow to manage systems topology through nesting and compartmentalisation, explicit spatial models are able to depict more precise localities and \emph{ecological niches}, describing how organisms or populations respond to the distribution of resources and competitors~\cite{LB98}. The spatial extensions of CWC~\cite{BCCDSST11}, CLS~\cite{BMMP11} and P-Systems~\cite{BMMPT11} could be used to express this kind of analysis allowing to deal with spatial coordinates.

\bibliographystyle{splncs03}
\bibliography{fmb}
\end{document}